\documentclass[aps,pra,preprint,superscriptaddress]{revtex4-1}
\usepackage{pstricks}
\usepackage{epsfig}
\usepackage{graphicx}
\usepackage{amsmath}
\usepackage{amsfonts}
\usepackage{amssymb}
\usepackage{wasysym}
\usepackage{bbold}
\usepackage{rotating}
\newcommand{\bra}[1]{\langle #1|}
\newcommand{\ket}[1]{|#1\rangle}

\begin{document}
\title{Variational density matrix optimization using semidefinite programming}
\author{Brecht Verstichel}
\email{brecht.verstichel@ugent.be}
\affiliation{Center for Molecular Modeling, Ghent University, Technologiepark 903, 9052 Zwijnaarde, Belgium}
\author{Helen van Aggelen}
\affiliation{Department of Inorganic and Physical Chemistry, Ghent University, Krijgslaan 281 (S3), 9000 Gent, Belgium}
\author{Dimitri Van Neck}
\affiliation{Center for Molecular Modeling, Ghent University, Technologiepark 903, 9052 Zwijnaarde, Belgium}
\author{Paul W. Ayers}
\affiliation{Department of Chemistry, McMaster University, Hamilton, Ontario, L8S 4M1, Canada }
\author{Patrick Bultinck}
\affiliation{Department of Inorganic and Physical Chemistry, Ghent University, Krijgslaan 281 (S3), 9000 Gent, Belgium}
\begin{abstract}
We discuss how semidefinite programming can be used to determine the second-order density matrix directly through a variational optimization. We show how the problem of characterizing a physical or $N$-representable density matrix leads to matrix-positivity constraints on the density matrix. We then formulate this in a standard semidefinite programming form, after which two interior point methods are discussed to solve the SDP. As an example we show the results of an application of the method on the isoelectronic series of Beryllium.
\end{abstract}
\maketitle
\section{Introduction}
The idea of a variational determination of  the ground-state energy for a non-relativistic many-body problem based
on the second-order density matrix (2DM) has a long history \cite{lowdin,coleman,garrod} and several highly appealing features.
The energy of a system is a known linear functional of the 2DM. $N$-particle wave functions never
need to be manipulated since the energy is minimized directly in terms of the 2DM.
However, the minimization is constrained because the  variational search should be done
exclusively with 2DMs that can be derived from an $N$-particle wave function (or an ensemble of $N$-particle wave functions).
Such a 2DM is called $N$-representable, and the complexity of the many-body problem is in fact shifted to the
characterization of this set of $N$-representable 2DMs.
The complete (necessary and sufficient) set of conditions for $N$-representability of a 2DM is not known in a constructive form, but it is clear that the energy from a minimization constrained by a set of necessary $N$-representability conditions is a strict lower bound to the exact energy. Therefore this approach is highly complementary to the usual variational procedure based on a wave-function ansatz, which produces upper bounds.
In addition the method is in principle exact, in the sense that as increasingly accurate set of $N$-representability conditions are imposed in the minimization, the resulting energy converges to the exact one.

These are fascinating ideas for any true-blooded many-body theorist, as it comes close to the ``ultimate reduction''
of an interacting many-particle problem to solving a sequence of two-particle problems. In practice, however,
implementing the method turns out to be very difficult and it is only in the last decade that serious attempts have been undertaken to turn the idea into a practical calculational scheme. The efforts by Mazziotti et al. \cite{mazziotti,hammond,maz_prl} and Nakata et al. \cite{nakata_first,nakata_last} are particularly notable. The main difficulty is of a technical nature: stringent $N$-representability conditions require the positive semidefiniteness of matrix functionals of the 2DM, which turns the variational problem into a so-called semidefinite program (SDP).
Even applying the simplest ``two-index'' conditions, a direct energy minimization using Newton-Raphson methods requires
a matrix operation scaling as $M^{12}$ (where $M$ is the number of single-particle states) in each Newton-Raphson step.
This can be circumvented in various ways, so that only matrix operations scaling as $M^6$ are needed.
While these are \textit{nominally} $M^6$ methods, the number of iterations required to reach convergence is
very high and seems to rise with system size; in practice present implementations are probably about 100-1000 times slower than comparable methods. Still, one has the feeling that there is potential to turn it into a \textit{genuine} $M^6$ method, and it is of interest to investigate the properties of SDP applied to various  systems.
In this proceeding we will first discuss the problem of density matrix optimization and $N$-representability, and how it can be formulated as an SDP. We will describe two different algorithms we used to solve the problem and as an example we will discuss the results of application of the method to the isoelectronic series of Be \cite{atomic}.
\section{Variational density matrix determination}
We use  second quantized notation where $a^\dagger_\alpha$ ($a_\alpha$) creates (annihilates) a fermion in a single-particle (sp) state $\alpha$). When there are only two-body interactions present in a physical system, the Hamiltonian of that system can be written as:
\begin{equation}
\hat{H} = \sum_{\alpha\gamma}t_{\alpha\gamma} a^\dagger_\alpha a_\gamma + \frac{1}{4}\sum_{\alpha\beta\gamma\delta}V_{\alpha\beta;\gamma\delta}a^\dagger_\alpha a^\dagger_\beta a_\delta a_\gamma~.
\end{equation}
The expectation value of the energy in an arbitrary $N$-particle state $\ket{\Psi^N}$ can be expressed using only the second-order density matrix (2DM),
\begin{equation}
E(\Gamma) = \mathrm{Tr}~\Gamma H^{(2)} = \frac{1}{4}\sum_{\alpha\beta\gamma\delta}\Gamma_{\alpha\beta;\gamma\delta}H^{(2)}_{\alpha\beta;\gamma\delta}~,
\label{ener_func}
\end{equation}
with the 2DM defined as:
\begin{equation}
\Gamma_{\alpha\beta;\gamma\delta} = \bra{\Psi^N}a^\dagger_\alpha a^\dagger_\beta a_\delta a_\gamma \ket{\Psi^N}~,
\label{2DM}
\end{equation}
and the reduced two-particle Hamiltonian,
\begin{equation}
H^{(2)}_{\alpha\beta;\gamma\delta} = \frac{1}{N-1}\left(\delta_{\alpha\gamma}t_{\beta\delta} - \delta_{\alpha\delta}t_{\beta\gamma} - \delta_{\beta\gamma}t_{\alpha\delta} + \delta_{\beta\delta}t_{\alpha\gamma}\right) + V_{\alpha\beta;\gamma\delta}~.
\end{equation}
Now the idea of variational density matrix optimization is to determine the ground-state energy and other two-body properties by minimizing the energy (\ref{ener_func}) using the 2DM as a variable. This is a much compacter object than the wavefunction because, no matter how many particles are involved, you always stay in two-particle (tp) space. The problem is that there is no straightforward way to know whether an arbitrary matrix in tp-space $\Gamma$ is derivable from a physical $N$-particle fermionic wavefunction as in Eq. (\ref{2DM}). This is called the $N$-representability problem. Some obvious necessary $N$-representability constraints are apparent from the definition (\ref{2DM}):
\begin{eqnarray}
\mathrm{Tr}~\Gamma &=& \frac{N(N-1)}{2}~,\\
\Gamma_{\alpha\beta;\gamma\delta} &=& -\Gamma_{\beta\alpha;\gamma\delta} = -\Gamma_{\alpha\beta;\delta\gamma} = \Gamma_{\beta\alpha;\delta\gamma}~,\\
\Gamma_{\alpha\beta;\gamma\delta} &=& \Gamma_{\gamma\delta;\alpha\beta}~,
\end{eqnarray}
but it turns out that there are many more non-trivial constraints needed to ensure that a 2DM is physical.
\subsection{$N$-representability}
The necessary and sufficient conditions for $N$-representability are known. A tp-matrix is $N$-representable if and only if, for every two-body Hamiltonian $\hat{H}_\nu$, the following inequality is satisfied:
\begin{equation}
\mathrm{Tr}~H^{(2)}_\nu \Gamma \geq E_0(\hat{H}_\nu)~.
\end{equation}
This is of course not a constraint that can be used in practice, as you need to know the ground-state energy of every imaginable two-body Hamiltonian. Therefore one resorts to certain classes of Hamiltonians for which a lower bound to the ground-state energy is known \cite{coleman,garrod,zhao}. A Hamiltonian class that is used as necessary constraint is
\begin{equation}
\label{stand_constr_tp}
\bra{\Psi^N}B^\dagger B\ket{\Psi^N} \geq 0~,
\end{equation}
which leads to positivity conditions of linear matrix maps of the 2DM. There are three possible forms of the operator $B^\dagger$ if we want (\ref{stand_constr_tp}) to be expressable as a function of the 2DM, which gives rise to three conditions on the density matrix:
\paragraph{$B^\dagger = \sum_{\alpha\beta}p_{\alpha\beta}a^\dagger_\alpha a^\dagger_\beta$} leads to the trivial $\mathcal{I}$-condition:
\begin{equation}
\sum_{\alpha\beta\gamma\delta}p_{\alpha\beta}\bra{\Psi^N}a^\dagger_\alpha a^\dagger_\beta a_\delta a_\gamma\ket{\Psi^N}p_{\gamma\delta}\geq 0\qquad\rightarrow\qquad\mathcal{I}(\Gamma) = \Gamma \succeq 0~,
\end{equation}
which just demands that the 2DM remains positive semidefinite.
\paragraph{$B^\dagger = \sum_{\alpha\beta}q_{\alpha\beta}a_\alpha a_\beta$} leads to the $\mathcal{Q}$-condition:
\begin{equation}
\sum_{\alpha\beta\gamma\delta}q_{\alpha\beta}\bra{\Psi^N}a_\alpha a_\beta a^\dagger_\delta a^\dagger_\gamma\ket{\Psi^N}q_{\gamma\delta}\geq 0\qquad\rightarrow\qquad\mathcal{Q}(\Gamma) \succeq 0~,
\end{equation}
in which the $\mathcal{Q}$ can be written as a function of $\Gamma$ using the anticommutation relations.
\paragraph{$B^\dagger = \sum_{\alpha\beta}g_{\alpha\beta}a^\dagger_\alpha a_\beta$} which leads to the $\mathcal{G}$-condition:
\begin{equation}
\sum_{\alpha\beta\gamma\delta}g_{\alpha\beta}\bra{\Psi^N}a^\dagger_\alpha a_\beta a^\dagger_\delta a_\gamma\ket{\Psi^N}g_{\gamma\delta}\geq 0\qquad\rightarrow\qquad\mathcal{G}(\Gamma) \succeq 0~,
\end{equation}
Another Hamiltonian class for which a lower bound to the ground-state energy is known, gives rise to the so-called three-index conditions:
\begin{equation}
\bra{\Psi^N}\left\{B^\dagger,B\right\}\ket{\Psi^N} \geq 0~.
\label{three_index}
\end{equation}
Two commonly used three-index conditions can be derived from Eq.~(\ref{three_index}):
\paragraph{$B^\dagger = \sum_{\alpha\beta\gamma}t^1_{\alpha\beta\gamma}a^\dagger_\alpha a^\dagger_\beta a^\dagger_\gamma$}
leads to the $\mathcal{T}_1$-condition:
\begin{equation}
\sum_{\alpha\beta\gamma\delta\epsilon\zeta}t^1_{\alpha\beta\gamma}\bra{\Psi^N}a^\dagger_\alpha a^\dagger_\beta a^\dagger_\gamma a_\zeta a_\epsilon a_\delta + a_\alpha a_\beta a_\gamma a^\dagger_\zeta a^\dagger_\epsilon a^\dagger_\delta\ket{\Psi^N}t^1_{\delta\epsilon\zeta}\geq 0\quad\rightarrow\quad\mathcal{T}_1(\Gamma) \succeq 0~.  \end{equation}
\paragraph{$B^\dagger = \sum_{\alpha\beta\gamma}t^2_{\alpha\beta\gamma}a^\dagger_\alpha a^\dagger_\beta a_\gamma$}
leads to the $\mathcal{T}_2$-condition
\begin{equation}
\sum_{\alpha\beta\gamma\delta\epsilon\zeta}t^2_{\alpha\beta\gamma}\bra{\Psi^N}a^\dagger_\alpha a^\dagger_\beta a_\gamma a^\dagger_\zeta a_\epsilon a_\delta + a^\dagger_\gamma a_\beta a_\alpha a^\dagger_\delta a^\dagger_\epsilon a_\zeta\ket{\Psi^N}t^2_{\delta\epsilon\zeta}\geq 0\quad\rightarrow\quad\mathcal{T}_2(\Gamma) \succeq 0~.
\end{equation}
In short, the optimization problem that we have to solve can be summerized as:
\begin{equation}
\min_{\Gamma} \mathrm{Tr}~\Gamma H^{(2)}~,
\end{equation}
under the condition that
\begin{eqnarray}
\mathrm{Tr}~\Gamma &=& \frac{N(N-1)}{2}~,\\
\mathcal{L}(\Gamma) &\succeq& 0 \qquad \forall \mathcal{L} \in \{\mathcal{I,Q,G},\mathcal{T}_1,\mathcal{T}_2\}~.
\end{eqnarray}
\section{Representation as an SDP}
The variational method described in the previous section can be formulated as a semidefinite program. A general 2DM, describing an $N$-particle system can be expanded in an arbitrary orthogonal basis of traceless matrix space $\{f^i\}$ as
\begin{equation}
\Gamma = \frac{N(N - 1)}{M(M-1)} \mathbb{1}_{\text{tp}} + \sum_i \gamma_i f^i~,
\end{equation}
with $M$ the dimension of single-particle (sp) space, and the unit matrix on tp space defined as
\begin{equation}
\left(\mathbb{1}_{\text{tp}}\right)_{\alpha\beta;\gamma\delta} = \delta_{\alpha\gamma}\delta_{\beta\delta} - \delta_{\alpha\delta}\delta_{\beta\gamma}~.
\end{equation}
The energy of the system can be written as a function of the $\gamma$'s as
\begin{equation}
\mathrm{Tr}~\Gamma H^{(2)} = \frac{N(N - 1)}{M(M-1)}\mathrm{Tr}~H^{(2)} + \sum_i \gamma_i \mathrm{Tr}~H^{(2)}f^i~.
\end{equation}
Because the necessary $N$-representability conditions can be written as linear homogeneous matrix maps of $\Gamma$, we can also write them as a function of the $\gamma$'s:
\begin{equation}
\mathcal{L}\left({\Gamma}\right) = \frac{N(N - 1)}{M(M-1)}\mathcal{L}\left({\mathbb{1}_\text{tp}}\right) + \sum_{i} \gamma_i \mathcal{L}\left({f^i}\right) \succeq 0~.
\end{equation}
If we now define the block matrices:
\begin{equation}
u^0 = \frac{N(N - 1)}{M(M-1)}\bigoplus_j \mathcal{L}_j\left(\mathbb{1}_\text{tp}\right) \qquad\text{and}\qquad u^i = \bigoplus_j \mathcal{L}_j\left(f^i\right)~,
\end{equation}
then we can formulate the variational density matrix optimization problem as a standard dual form of a semidefinite program \cite{boyd}:
\begin{equation}
\min_\gamma~\gamma^T h \qquad \text{on condition that} \qquad Z = u^0 + \sum_i \gamma_i u^i \succeq 0~,
\label{dual}
\end{equation}
in which $h^i = \mathrm{Tr}~H^{(2)}f^i$. The primal problem corresponding to (\ref{dual}) optimizes the matrixvariable $X$, the problem is defined as:
\begin{equation}
\max_X~ -\mathrm{Tr}~Xu^0 \qquad \text{on condition that} \qquad \mathrm{Tr}~Xu^i = h^i \qquad \text{and} \qquad X\succeq 0~.
\label{primal}
\end{equation}
The primal-dual gap $\eta$ is defined as the difference between the primal and the dual cost function for a certain primal-dual point $(X,Z)$:
\begin{equation}
\eta = \gamma^T h + \mathrm{Tr}~u^0 X = \sum_i \gamma_i \mathrm{Tr}~X u^i + \mathrm{Tr}~X u^0 = \mathrm{Tr}~X Z \geq 0~,
\end{equation}
because $X$ and $Z$ are positive semidefinite matrices. We can see that the smallest value of $\eta$ will be reached when both the primal and the dual problem are optimal. It can be proved \cite{boyd} that if the primal and the dual problem are both feasible, then at their solution the primal-dual gap will vanish. This means that the primal-dual gap can be used as a convergence criterion for the algorithm, what is more, at any point during the optimization, the error on the current value is limited from above by the primal-dual gap.
\section{Algorithms used to solve SDP}
There is a vast literature available with different methods to solve SDP's. It is generally excepted that the interior point methods have the best computational performance. In the course of our research we have implemented numerous algorithms, here we will discuss two of them, a dual-only potential reduction method, and a primal-dual path following method.
\subsection{Dual potential reduction method}
This method will only solve the dual problem (\ref{dual}), so we have no access to the primal-dual gap as a convergence criterion. The idea is to minimize the following potential function over $\gamma$:
\begin{equation}
\label{pot_red}
\Phi(\gamma,t) = \gamma^T h - t\ln{\det Z(\gamma)}~.
\end{equation}
Starting from a feasible $Z$, the logarithmic potential will make sure that $Z$ remains positive definite. The potential is minimized for decreasing scaling factor $t$, using the solution of the previous $t$ as a starting point. When $t \rightarrow 0$ the solution will lie on the edge of the feasible area, and this is exactly the solution of the original semidefinite program (\ref{dual}). For every value of $t$ the optimization problem of (\ref{pot_red}) will be solved using Newton method. This means that we have to solve the following linear system of equations to find the step $\delta\gamma$ until convergence is reached:
\begin{equation}
\label{Newton_1}
t\sum_j\left(\mathrm{Tr}~u^iZ^{-1}u^jZ^{-1}\right)(\delta\gamma)_j = t\mathrm{Tr}Z^{-1}u^i - h^i~.
\end{equation}
The dimension of the system of equations is $M^4$, which means that if we would solve it by direct inversion the method would scale as $M^{12}$. We can construct an efficient matrix-vector product and use the conjugate gradient method to solve this much faster. Unfortunately in the limit of small $t$ the number of iterations needed to converge increases because the system becomes ill conditioned.
\subsection{Primal-dual path following method}
In this method we will solve the primal and the dual problem at the same time. To explain how it works we first have to define the central path, which is the set of primal-dual points for which
\begin{equation}
\label{cent_path}
X Z = \frac{\eta}{n} \mathbb{1}_\text{sup}~,
\end{equation}
with $n$ the total dimension of the $X$ and $Z$ matrices and $\mathbb{1}_\text{sup}$ the direct sum of the unity matrices on the different constraint spaces:
\begin{equation}
\mathbb{1}_\text{sup} = \bigoplus_k \mathbb{1}_k~.
\end{equation}
In the path following algorithm we will try to follow the central path, reducing the primal-dual gap along the way. Imagine that we have a primal-dual point $(X,Z)$ on the central path with primal-dual gap $\eta$. We want to answer the question: what is the primal-dual point on the central path with primal-dual gap scaled down with a factor $\nu$, rephrased, what are the $(\Delta X,\Delta Z)$ that solve:
\begin{equation}
(X + \Delta X)(Z + \Delta Z) = \frac{\nu\eta}{n}\mathbb{1}_\text{sup}~.
\label{EOM}
\end{equation}
There are several ways to symmetrize these equations, using the method proposed by Nesterov and Todd (see \cite{nesterov}), we obtain two equivalent equations, which we call the primal and the dual equation:
\begin{eqnarray}
\label{P_eom}(P) : \Delta Z + D^{-1}\Delta X D^{-1} &=& \frac{\nu\eta}{n} X^{-1} - Z~,\\
\label{D_eom}(D) : \Delta X + D~\Delta {Z}~D &=& \frac{\nu\eta}{n} Z^{-1} - X~,
\end{eqnarray}
with 
\begin{equation}
D(X,Z) = Z^{-\frac{1}{2}}\left(Z^{\frac{1}{2}}XZ^{\frac{1}{2}}\right)^{\frac{1}{2}}Z^{-\frac{1}{2}}~,
\label{metric}
\end{equation}
and under the condition that:
\begin{equation}
\label{eom_constr}
\mathrm{Tr}~\Delta Xu^i = 0 \qquad \text{and} \qquad \Delta Z = \sum_i (\delta \gamma)_i u^i~.
\end{equation}
We can see that now, we will have to solve two systems of linear equations:
\begin{eqnarray*}
\sum_j \left(\mathrm{Tr}~D~c^j~D~c^i\right) \delta z_j = \mathrm{Tr}~\left(\frac{\nu\eta}{n} X^{-1} - Z~\right) c^i~,\\
\sum_j \left(\mathrm{Tr}~D^{-1} u^j D^{-1} u^i\right) (\delta\gamma)_j = \mathrm{Tr}~\left(\frac{\nu\eta}{n} Z^{-1} - X\right) u^i~,
\end{eqnarray*}
in which we define $\{c^i\}$ as the basis of the orthogonal complement of the space spanned by the $u^i$'s, and $\Delta Z = \sum_j \delta z_j c^j$. We will again be able to construct an efficient matrix-vector product and use the conjugate gradient method. First we solve the dual system, which has the smallest dimension, and feed the solution into the primal system, which then needs very few iterations to converge. Once again we run into the same problem that the solution of the dual system needs more and more iterations as it gets closer to the solution, because of the increasing condition number of the system.
\section{Application to Beryllium isoelectronic series}
We have applied this method in the study of electronic structure calculations of diatomic molecules \cite{helen_1,helen_2,qsep} and atomic systems \cite{atomic}. We will discuss in short the result of the method applied to the isoelectronic series of Beryllium, using the $P$, $Q$ and $G$ conditions. This is an interesting test case for the inclusion of static electron correlation, because there is a near degeneracy between the $2s$ and $2p$ energy levels when the central charge becomes large, thus creating a multireference ground state. Because of the amount of symmetry present in atomic systems, the density matrix and the constraint matrices all become block diagonal (see \cite{atomic}), enabling us to use quite large basis sets (cc-pVDZ, cc-pVTZ and cc-pVQZ \cite{dunning}).
\begin{figure}
\includegraphics[scale=0.6]{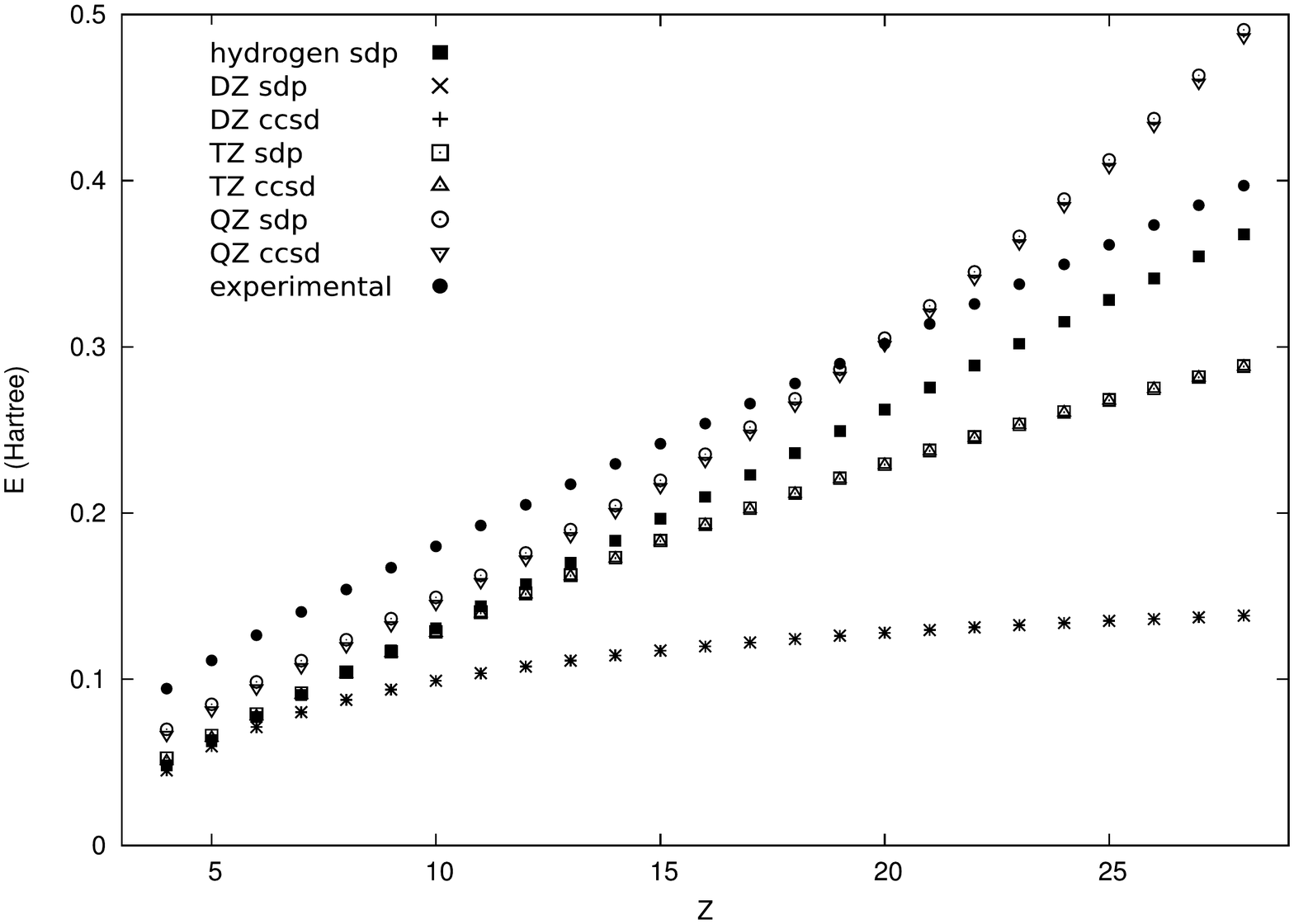}
\caption{\label{beryl_corr_ener} The SDP correlation energy ($E_{\text{HF}} - E_{\text{SDP}}$) for the Be series in all three basis sets, and in a rescaled basis set that exhibits hydrogen-like behavior (degeneracy between the $2s$ and $2p$ level). For comparison, the CCSD and experimental values are also shown.}
\end{figure}
In Fig.~\ref{beryl_corr_ener} the SDP correlation energy, defined as the difference between the Hartree-Fock and the SDP energy, is shown as a function of central charge $Z$ for the different basis sets, compared to estimates for non-relativistic energies based on experimental data \cite{davidson,chakravorty}, and to the results of coupled cluster (CCSD) calculations. The experimental curve is linear in $Z$, as a direct consequence of the near-degeneracy of the ground state \cite{chakravorty}. The SDP correlation energy does not follow this trend: it goes linear in the beginning, but becomes concave in the cc-pVDZ and cc-pVTZ basis, or convex in the cc-pVQZ basis. This failure, however, is not related to the SDP method as the trend is the same in the CCSD calculations. It simply reflects the fact that the incipient degeneracy is not well described in these basis sets. This can also be seen by calculating the hydrogen spectrum (corresponding to the $Z\rightarrow \infty$ situation, when the electron-electron interaction can be neglected) in the basis sets: the $2s$ and $2p$ energies are not degenerate, but differ by 5.8 mhartree (cc-pVDZ), 2.0 mhartree (cc-pVTZ) and -2.3 mhartree (cc-pVQZ). We also performed calculations in the cc-pVDZ basis after rescaling  ($r\rightarrow\alpha r$) it in such a way that the hydrogenic $2s$-$2p$ degeneracy is exact. In this basis the SDP correlation energy (also shown in Fig.~\ref{beryl_corr_ener}) indeed has the correct linear behavior. It is clear from the above discussion that SDP is indeed capable of providing accurate correlation energies in the presence of near degeneracies, when other many-body techniques (like density functional theory or MP2) can fail.
\bibliography{ccp2010.bib}
\bibliographystyle{unsrt}
\end{document}